\documentclass[12pt]{article}
\usepackage{newtxtext,newtxmath}

\usepackage{graphicx}
\usepackage{caption}
\usepackage{xcolor}
\usepackage{makecell}

\usepackage[letterpaper,margin=1in]{geometry}

\renewenvironment{abstract}
	{\quotation}
	{\endquotation}

\date{}

\makeatletter
\renewcommand{\fnum@figure}{\textbf{Fig. \thefigure . }}
\renewcommand{\fnum@table}{\textbf{Table \thetable .}}
\newcommand{\angstrom}{\ensuremath{\mathring{\text{A}}}}
\makeatother

\usepackage{scicite}
\usepackage[T1]{fontenc} 

\usepackage{url}

\def\scititle{Floquet optical selection rules in black phosphorus}
\title{\bfseries \boldmath \scititle}

\author{
	Benshu~Fan$^{1,2}$,
	Umberto~De~Giovannini$^{2,3}$,
	Hannes~H\"ubener$^{2}$,
        Shuyun Zhou$^{1,4}$,\and
        Wenhui Duan$^{1,5,4\ast}$,
        Angel~Rubio$^{2,6\ast}$,
        Peizhe Tang$^{7,2\ast}$\and
    \small$^{1}$State Key Laboratory of Low-Dimensional Quantum Physics and Department of Physics,\and
    \small Tsinghua University, Beijing 100084, People's Republic of China.\and
    \small$^{2}$Max Planck Institute for the Structure and Dynamics of Matter, \and
    \small Center for Free Electron Laser Science, 22761 Hamburg, Germany.\and
    \small$^{3}$Università degli Studi di Palermo, Dipartimento di Fisica e ChimicaEmilio Segrè, \and
    \small via Archirafi 36, I-90123 Palermo, Italy.\and
    \small$^{4}$Frontier Science Center for Quantum Information, Beijing 100084, People's Republic of China.\and
    \small$^{5}$Institute for Advanced Study, Tsinghua University, Beijing 100084, People's Republic of China.\and
    \small$^{6}$Center for Computational Quantum Physics (CCQ),\and
    \small The Flatiron Institute, 162 Fifth avenue, New York NY 10010, USA.\and
    \small$^{7}$School of Materials Science and Engineering, Beihang University, Beijing 100191, People's Republic of China.\and
	\small$^\ast$Corresponding author. Email: duanw@mail.tsinghua.edu.cn (W.D.); \and
    \small angel.rubio@mpsd.mpg.de (A.R.); peizhet@buaa.edu.cn (P.T.)\and
}

\begin{document} 
\captionsetup[figure]{labelsep=none}

\maketitle

\begin{abstract} \bfseries \boldmath
Optical selection rules endorsed by symmetry are crucial for understanding the optical properties of quantum materials and the associated ultrafast spectral phenomena. Here, we introduce momentum-resolved Floquet optical selection rules using group theory to elucidate the pump-probe photoemission spectral distributions of monolayer black phosphorus (BP), which are governed by the symmetries of both the material and the lasers. Using time-dependent density functional theory (TDDFT), we further investigate the dynamical evolution of Floquet(-Volkov) states in the photoemission spectra of monolayer BP, revealing their spectral weights at specific momenta for each sideband. These observations are comprehensively explained by the proposed Floquet optical selection rules. Our framework not only clarifies experimental photoemission spectra but also uncovers unexplored characteristics under different pump-probe configurations. Our results are expected to deepen the understanding of light-induced ultrafast spectra in BP and can be extended to other Floquet systems.
\end{abstract}

\subsection*{Teaser}
Symmetry-based Floquet optical selection rules determine the visibility of light-induced sidebands in TrARPES intensity plots.

\subsection*{Introduction}
Ultrafast lasers have emerged as indispensable tools for detecting and manipulating optical and electronic properties of quantum materials, enabling the exploration of light-matter interaction through symmetry-based paradigms \cite{HsiehDemond2017,DelaTorre2021}. A representative example lies in selection rules for detecting related optical properties in equilibrium, which dictate allowable electric dipole transitions under different linearly \cite{qiao2014high,yuan2015polarization} or circularly \cite{mak2012control,zeng2012valley} polarized lasers governed by the point group symmetry of the target material. Beyond probing equilibrium characterization, the light field can also drive the system into a non-equilibrium regime, leading to distinctive physical phenomena \cite{bloch2022strongly,bao2022light}. For instance, via virtual absorption or emission of multiple photons, the Bloch states can be dressed into Floquet states inside the crystal and Volkov states in the vacuum by the strong time-periodic field \cite{WangYH2013,Aeschlimann2021Survival,Lee2021Steady,ito2023build}, and their interference can further give rise to the Floquet-Volkov states \cite{Park2014Interference,mahmood2016selective,choi2025observation,merboldt2024observation}. Moreover, accompanied by the formation of light-induced sidebands in the energy domain, the light field can transiently modify electronic \cite{oka2009photovoltaic,Lindner2020,lindner2011floquet,sentef2015theory,chan2016chiral,yan2016tunable,claassen2016all,de2016monitoring,hubener2017creating,liu2018photoinduced,sato2019microscopic,mciver2020light,neufeld2023band,zhu2023floquet,liu2023floquet,fanbs2024chiral} and optical properties \cite{sie2015valley,sie2017large,Hsieh2021giant,zhang2024light} of target materials, known as the Floquet engineering \cite{oka2019floquet,bao2022light}.

Recently, time- and angle-resolved photoemission spectroscopy (TrARPES) experiments have provided the first observation of the Floquet band renormalization in the semiconductor BP thin film driven by the linearly polarized pumping laser \cite{zhou2023pseudospin,zhou2023Floquet,bao2024light}. Intriguingly, such Floquet engineering occurs for the pumping laser polarized along the armchair (AC) direction but not the zigzag (ZZ) direction, exhibiting a strong pump polarization dependence. This phenomenon underscores the point group symmetry-enforced pseudospin selectivity in the Floquet band engineering for BP. In this context, the pseudospin is a generalized concept of spin in a two-sublattice system with distinct behaviors from spin under spatial symmetry operations \cite{castro2009electronic,chung2024dark}, it governs optical selection rules in equilibrium, which refer to symmetry-imposed constraints on the electric dipole transitions induced by an external light field, as confirmed by angle-resolved photoemission spectroscopy (ARPES) experiments \cite{jung2020black}, and notably influences the Floquet band structures \cite{zhou2023pseudospin,zhou2023Floquet,bao2024light}.

Despite prior advances in Floquet engineering on band structures, recent experimental findings on BP \cite{zhou2023Floquet,SYZhou2024spot} reveal symmetry-sensitive TrARPES signatures under certain pump-probe configurations that cannot be fully explained by existing theoretical models. For example, when both pumping and probe lasers are polarized along the AC direction, the TrARPES intensity plot shows missing spectral weights around the $\Gamma$ point for the $n$=1 Floquet sideband, differing from the $n$=0 Floquet sideband features \cite{zhou2023Floquet}. Moreover, when the probe laser polarization switches to the ZZ direction while maintaining the AC pumping configuration, only a spot-like spectral signature appears around the $\Gamma$ point for the $n$=1 Floquet sideband \cite{SYZhou2024spot}. These phenomena demonstrate the strong dependence of the light-induced sideband visibility on the light polarization, but without a comprehensive understanding from a general theoretical framework. 

On the other hand, while some dynamical selection rules have been proposed for the Floquet systems, they primarily focus on the high harmonic generation to integrate all contributions from each $k$ point \cite{neufeld2019floquet,tzur2022selection,lerner2023multiscale}, or are limited to molecular and model systems without fully accounting for the point group symmetries \cite{wang2021observation,engelhardt2021dynamical}. As such, existing frameworks do not capture the momentum-resolved selection rules necessary to interpret experimental TrARPES features in materials like BP and generally leave the detectability of light-induced sidebands unexplained and unpredictable. Consequently, there is a critical need to propose general symmetry-based Floquet optical selection rules, which could depict the photoemission process between the photon-dressed
sideband states and vacuum states under various pump-probe configurations. These rules are essential for interpreting and predicting the spectral features of sidebands in real materials and are crucial tools for the future of Floquet engineering.

Motivated by these open questions, in this work, we choose monolayer BP, which shares the same point group symmetry with the BP thin film, as a showcase system to investigate Floquet optical selection rules under various pump-probe configurations. These rules jointly depend on the intrinsic lattice symmetry and the polarization geometries of the pumping and probe lasers. We also employ a state-of-the-art \emph{ab initio} calculation approach to simulate the photoemission spectra, which is the time-dependent surface flux method in the semi-periodic system (t-SURFFP) \cite{tao2012photo,de2017first} based on real-time TDDFT (see Methods for more details). In this approach, the ionization process is simulated by allowing electrons to escape the sample under the action of the probe field and be collected by a "numerical" detector placed at a fixed distance from the surface. This enables the computation of high-quality (Tr)ARPES intensity plots that outperform simpler approximations, especially in capturing surface-specific effects and time-resolved dynamics \cite{de2020direct,kern2023simple}. Besides capturing the build-up, evolution, and collapse of light-induced sidebands to confirm the Floquet engineering, our spectral simulations also track the real-time evolution of electron occupations in these Floquet(-Volkov) sidebands, thereby reflecting their controllable features under various pump-probe conditions. Importantly, in addition to reproducing previously reported spectral features \cite{zhou2023Floquet,SYZhou2024spot}, our TDDFT simulations uncover spectral weight distributions that have not been reported in earlier works. These findings provide predictive insights into the symmetry-governed visibility or suppression of sidebands, which are fully accounted for by our proposed Floquet optical selection rules. Taken together, our work offers a general theoretical framework for understanding and predicting symmetry-governed features in TrARPES, which can be further extended to other light-driven quantum materials beyond BP.

\subsection*{Results}
\begin{figure}[h]
	\centerline{\includegraphics[clip,width=0.8\linewidth]{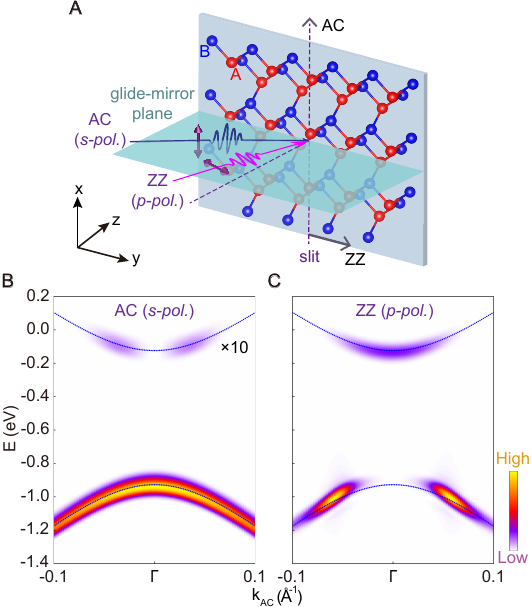}}
	\caption{\textbf{The simulated ARPES intensity plots for probe lasers polarized along different directions.} (\textbf{A}) Schematic for ARPES on monolayer BP with different probe polarizations. (\textbf{B}) The ARPES intensity plot along the AC direction for the $n$-doped monolayer BP under the AC ($s$-$pol.$) probe laser. The ARPES signature on the CB in (B) is magnified by a factor of 10 for clearer presentation. (\textbf{C}) Similar results to (B) but under the ZZ ($p$-$pol.$) probe laser. The band structures for both panels are in blue dot lines. The probe photon energy $\hbar\omega$ is 6.2 eV, and the peak intensity $I_{pr}$ is 0.03$\times10^9$ W/cm$^2$ with a pulse duration $T_{pr}$ of 80 fs.}
	\label{fig:1}
\end{figure}

Herein, considering the wealth of TrARPES experimental results on the BP thin film \cite{zhou2023pseudospin,zhou2023Floquet,SYZhou2024spot,bao2024light}, we choose the monolayer BP as a prototype material, which is a direct gap semiconductor at the $\Gamma$ point, featuring anisotropic electronic structures along the AC and ZZ directions \cite{ezawa2014topological,tran2014layer,rodin2014strain}. Under the glide-mirror operation with the green plane in Fig.~\ref{fig:1}A, which serves as the scattering plane in the pump-probe photoemission measurements \cite{jung2020black,zhou2023Floquet,zhou2023pseudospin,bao2024light,SYZhou2024spot}, the BP remains invariant. At the $\Gamma$ point, the valence band (VB) and conduction band (CB) wavefunctions arise from the superposition of wavefunctions $|A\rangle$ and $|B\rangle$ at A and B sublattices in Fig.~\ref{fig:1}A, namely $|A\rangle-|B\rangle$ and $|A\rangle+|B\rangle$. These two wavefunctions transform as the irreducible representations $\Gamma_2^+$ and $\Gamma_4^-$ of $D_{2h}$ group, with characters of $-1$ and 1 respect to the mirror operation $M_{yz}$ \cite{Kim2017prl}. In analogy to graphene, this sublattice-based structure gives rise to a well-defined pseudospin degree of freedom \cite{jung2020black}, with opposite pseudospin polarizations in the VB and CB. The pseudospin governs the polarization dependence of optical transitions and photoemission spectral features, and its role is consistent with previous findings in BP thin films \cite{zhou2023pseudospin}, as further detailed in Note S1 of the Supplementary Materials (SM).

To reveal optical selection rules in equilibrium as a benchmark, we conduct TDDFT simulations of ARPES intensity plots along the AC direction for $n$-doped monolayer BP. As illustrated in Fig.~\ref{fig:1}A, once we set the glide-mirror plane as the scattering plane in pump-probe photoemission measurements \cite{jung2020black,zhou2023Floquet,zhou2023pseudospin,SYZhou2024spot,bao2024light}, under $s$-polarization ($s$-$pol.$) probe laser with the electric field oscillating along the AC direction [noted as AC ($s$-$pol.$) probe laser], a distinct ARPES intensity signature emerges around the $\Gamma$ point for the VB in Fig.~\ref{fig:1}B. In contrast, almost no ARPES intensity signature is observed for the CB around the $\Gamma$ point (see Fig.~\ref{fig:1}B). Conversely, when the probe laser is altered to $p$-polarization ($p$-$pol.$) with the electric field oscillating in the scattering plane [noted as ZZ ($p$-$pol.$) probe laser, see Fig.~\ref{fig:1}A], the ARPES intensity signature around the $\Gamma$ point exhibits the opposite behavior: the VB vanishes entirely while the CB becomes distinctly visible, as shown in Fig.~\ref{fig:1}C. Our simulated spectra align qualitatively with previous ARPES experiments on BP thin film with electron doping \cite{jung2020black,zhou2023pseudospin}, demonstrating that electrons in the VB and CB can be selectively excited by tuning the polarization direction of the laser. This behavior reflects the underlying optical selection rules in BP.

\begin{figure}[h]
	\centerline{\includegraphics[clip,width=0.75\linewidth]{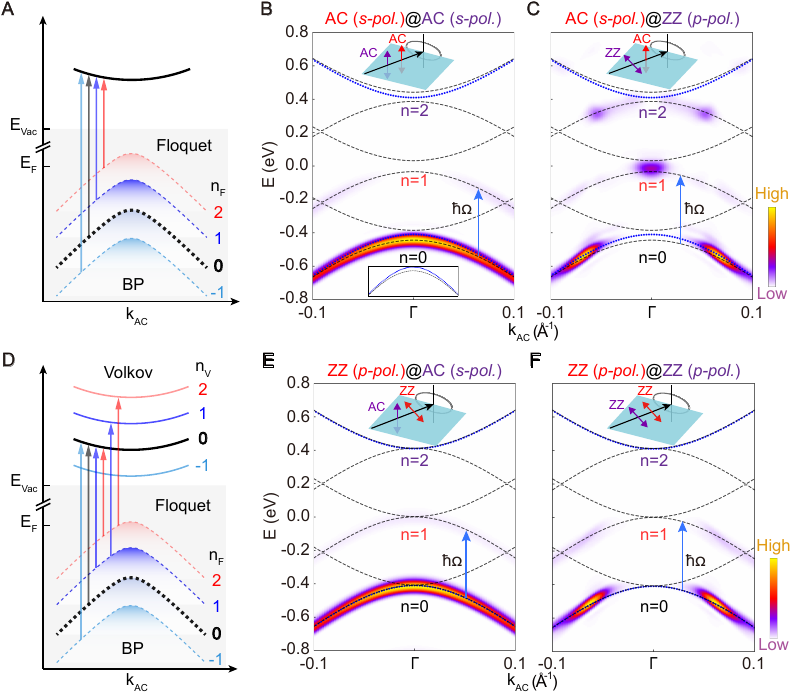}}
	\caption{\textbf{The simulated TrARPES intensity plots upon below-gap pumping.} (\textbf{A}) Schematic of Floquet states. (\textbf{B}) The TrARPES intensity plot along the AC direction upon below-gap pumping. The polarization configuration shown in the top inset is denoted as "pump@probe", with the pumping laser in red and the probe laser in purple. For example, "AC ($s$-$pol.$)@AC ($s$-$pol.$)" indicates that both lasers are polarized along the AC ($s$-$pol.$) directions. For the pumping laser, the photon energy $\hbar\Omega$ is 0.41 eV, which is half of the equilibrium band gap of 0.82 eV, and the peak intensity $I_{pu}$ is $4\times10^9$ W/cm$^2$ with a pulse duration $T_{pu}$ of 100 fs. The parameters of the probe laser remain consistent with the ARPES calculations but with a longer pulse duration $T_{pr}$ of 100 fs. The pump-probe delay time $\Delta t$ is set to 0 fs. The blue arrow denotes the single-photon absorption and the bottom inset shows the VB in blue dot lines for improved clarity. (\textbf{C}) Similar results with the same parameters as (B), but the polarization of the probe laser is along the ZZ ($p$-$pol.$) direction. (\textbf{D}) Schematic of Floquet-Volkov coherence. (\textbf{E} and \textbf{F}) Corresponding results with the same parameters as (B) and (C) but upon ZZ ($p$-$pol.$) pumping laser. The blue dot lines are equilibrium band structures and the black dashed lines are light-induced band structures with the Floquet index $n$ as $\{-2,-1,0,1,2\}$ for these photoemission spectra. The arrows in (A) and (D) denote the possible electronic transitions induced by the probe laser.}
	\label{fig:2}
\end{figure}

We further employ TDDFT to explore the formation of the Floquet(-Volkov) states in the undoped monolayer BP under laser pumping. The absence of electron doping offers a clean platform for the unambiguous resolution of the spectral weights of each sideband for the VB across different pump-probe conditions. Initially, we concentrate on the AC ($s$-$pol.$) pumping laser without an out-of-plane ($z$ direction) electric field component. This ensures no coupling between the light field and the photoemission final states around the $\Gamma$ point, forming the Floquet states. Moreover, the below-gap pumping with photon energy smaller than the band gap guarantees no single-photon absorption to excite electrons from the VB to CB. Consequently, we only focus on the VB edge and its light-induced replica bands with the Floquet index $n_{\rm{F}}$ (see Fig.~\ref{fig:2}A). Employing the Floquet TDDFT approach (see Methods), we simulate the Floquet band structures, labeled as $n=n_{\rm{F}}$ and illustrated by black dashed lines in Fig.~\ref{fig:2} (B and C). Compared with the equilibrium band structures in blue dot lines, we find a momentum-dependent energy shift around the $\Gamma$ point for the $n$=0 Floquet sideband for the VB (Floquet VB). This result is consistent with experimental observation under the same setup and light-pumping \cite{zhou2023Floquet}, underscoring that the Floquet TDDFT approach can effectively capture all features of the Floquet engineering on band structures.

We also calculate TrARPES intensity with the probe laser along different polarization directions to further reveal spectral features. TrARPES intensity plots shown in Fig.~\ref{fig:2} (B and C) exhibit similar features to those for equilibrium VB in Fig.~\ref{fig:1} (B and C). The $n$=0 Floquet VB around the $\Gamma$ point is visible under the AC ($s$-$pol.$) probe laser but not under the ZZ ($p$-$pol.$) probe laser. These results indicate that the $n=0$ Floquet VB inherits the symmetry of the equilibrium VB. Interestingly, the $n=1$ Floquet VB around the $\Gamma$ point exhibits distinct symmetry behavior. As shown in Fig.~\ref{fig:2}B, it is undetectable under the AC probe laser \cite{zhou2023Floquet}, while under the ZZ probe laser (see Fig.~\ref{fig:2}C), a spot-like spectral signature appears \cite{SYZhou2024spot}. These observations underscore that TrARPES spectral weights strongly depend on the symmetry of equilibrium VB and the polarization geometry of the probe laser.

In addition to the Floquet states in monolayer BP, the ZZ ($p$-$pol.$) pumping laser has a nonzero out-of-plane component of the electric field, which can dress the photoemission final states in the vacuum as the Volkov states with the Volkov index $n_{\rm{V}}$ (see Fig.~\ref{fig:2}D). Due to the Floquet-Volkov interference \cite{mahmood2016selective}, the $n$-th sideband in TrARPES intensity plots stems from transitions where $n=n_{\rm{V}}+n_{\rm{F}}$, as shown in Fig.~\ref{fig:2}D. The simulated results under the ZZ ($p$-$pol.$) pumping laser are presented in Fig.~\ref{fig:2} (E and F) with different probe lasers, showcasing no band renormalization for the $n$=0 VB. This finding is consistent with previous experimental and theoretical studies on BP thin film \cite{zhou2023Floquet}. For the experimental setup depicted in the inset of Fig.~\ref{fig:2}E, we calculate the dynamical evolution for the Floquet-Volkov states and the transfer of spectral weights among different light-induced sidebands by varying the pump-probe delay time $\Delta t$ (see Note S3.1 of SM). Our TDDFT simulations successfully visualize the formation and collapse of the Floquet-Volkov states in the time domain. Furthermore, the spectral weight distributions in each light-induced sideband at $\Delta t$=0 under AC ($s$-$pol.$) and ZZ ($p$-$pol.$) probe lasers also differ from those of Floquet states shown in Fig.~\ref{fig:2} (B and C). The $n$=0,1 sidebands for the VB around the $\Gamma$ point are visible under the AC ($s$-$pol.$) probe laser (see Fig.~\ref{fig:2}E), but disappear entirely under the ZZ ($p$-$pol.$) probe laser (see Fig.~\ref{fig:2}F), which has not been reported in previous works. Thus, based on these systemic simulations, we conclude that spectral weights for the Floquet(-Volkov) sidebands in monolayer BP depend on the Floquet index $n$ and symmetries of targeted bands in equilibrium, pumping lasers, and probe lasers. Although previous experimental studies have partially revealed polarization-dependent sideband intensities under specific pump-probe configurations \cite{SYZhou2024spot}, a unified and symmetry-based theoretical framework has remained elusive.

To address this, we aim to understand how symmetry governs the spectral features observed in (Tr)ARPES experiments. Given that (Tr)ARPES intensity is proportional to the square of the matrix element $\mathcal{M}$, which inherently encodes the symmetry details of both the experimental geometry and the wavefunction of the targeted state \cite{damascelli2003angle}, we apply the group theory to explore its property and propose comprehensive optical selection rules for the Floquet(-Volkov) states and equilibrium states. $\mathcal{M}$ can be expressed as 
\begin{equation}
\label{eq:matrix_element}
\mathcal{M}=\langle\psi_{f}|\hat{H}^\prime|\psi_{i}\rangle
\end{equation}
Herein, $\left|\psi_i\right\rangle$ and $\left|\psi_f\right\rangle$ are the wavefunctions of the initial and final states in (Tr)ARPES measurements. The light-matter interaction Hamiltonian $\hat{H}^\prime\simeq\mathbf{A}_{pr}\cdot\mathbf{\hat{p}}$ involves the vector potential of the probe laser $\mathbf{A}_{pr}$ and the electron momentum operator $\hat{\mathbf{p}}$. Given the scattering geometry shown in Fig.~\ref{fig:1}A, the final state $\left|\psi_f\right\rangle$ should have even symmetry under the glide-mirror operation \cite{damascelli2003angle}. To yield a non-zero photoemission intensity, $\mathcal{M}$ must be an even function under the mirror operation $M_{yz}$, implying that $\hat{H}^\prime|\psi_{i}\rangle$ in Eq.~\ref{eq:matrix_element} is also even. If $\hat{H}^\prime|\psi_{i}\rangle$ is an odd function, $\mathcal{M}$ is zero, and the target state is invisible in the photoemission spectrum (see Note S2.1 of SM). 

As a reference, we analyze the ARPES intensity plots for monolayer BP without laser pumping, as shown in Fig. \ref{fig:1} (B and C). The initial state in the matrix element $\mathcal{M}$ is the VB maximum (CB minimum), described by the wavefunction $\left|\psi^v\right\rangle$ ($\left|\psi^c\right\rangle$). At the $\Gamma$ point, the wave vector group is $D_{2h}$, under which the VB (CB) wavefunction $\left|\psi^v\right\rangle$ ($\left|\psi^c\right\rangle$) is odd (even) with respect to the $M_{yz}$ operation \cite{jung2020black}. Moreover, the AC ($s$-$pol.$) and ZZ ($p$-$pol.$) probe lasers, polarized perpendicular and parallel to the scattering plane, are odd and even with respect to $M_{yz}$, respectively. As a result, $\hat{H}^\prime|\psi^v\rangle$ ($\hat{H}^\prime|\psi^c\rangle$) is even under the $M_{yz}$ operation, indicating that the matrix element $\mathcal{M}$ is nonzero and the AC (ZZ) probe laser can selectively detect the signature of the VB (CB) edge at the $\Gamma$ point (see details in Note S2.2 of SM). These symmetry-based selection rules explain the opposite ARPES intensities in Fig.~\ref{fig:1} (B and C).

We further develop the Floquet optical selection rules for monolayer BP under laser pumping within the same framework. In this case, the initial states in the matrix element $\mathcal{M}$ correspond to the wavefunctions of light-induced sidebands. Considering that the pump photon energy is half the band gap, we focus on the VB and disregard the light-induced hybridization between the sidebands. To describe the VB edge around the $\Gamma$ point for monolayer BP, we adopt a parabolic single-band Hamiltonian $\hat{H}_\Gamma(\mathbf{k}_{\|})$, where $\mathbf{k}_{\|}$ denotes the in-plane momentum. The influence of the pumping laser is incorporated via the Peierls substitution, leading to a time-dependent effective Hamiltonian $\hat{H}_\Gamma(t,\mathbf{k}_{\|})=\hat{H}_\Gamma(\mathbf{k}_{\|})+\mathbf{v(\mathbf{k}_{\|})}\cdot\mathbf{A}_{pu}\cos\Omega t$, where $\mathbf{v(\mathbf{k}_{\|})}$ represents the in-plane velocity and $\mathbf{A}_{pu}$ is the vector potential of the pumping laser with the frequency of $\Omega$. By applying the Floquet theory \cite{oka2019floquet}, we derive the wavefunction of the $n$-th Floquet VB at the $\Gamma$ point $\left|\psi_n^v\right\rangle$, which acts as the initial state $\left|\psi_i\right\rangle$ in Eq.~\ref{eq:matrix_element} (see Note S2.3 of SM)
\begin{equation}
\label{eq:sideband}
\left|\psi_i\right\rangle=\left|\psi_n^v\right\rangle\propto(\mathbf{v(\mathbf{k}_{\|})}\cdot\mathbf{A}_{pu})^{|n|}\left|\psi^v\right\rangle
\end{equation}
For the AC ($s$-$pol.$) pumping and AC ($s$-$pol.$) probe lasers (see Fig.~\ref{fig:2}B), the representation of $\hat{H}^\prime|\psi_{i}\rangle$ at the $\Gamma$ point in Eq.~\ref{eq:matrix_element} is
\begin{equation}
\label{eq:AC_AC}
\underbrace{\Gamma_3^-}_{probe}\otimes(\underbrace{\Gamma_3^-}_{pump})^{|n|}\otimes\underbrace{\Gamma_2^+}_{\left|\psi^v\right\rangle}
\end{equation}
Since $\Gamma_3^-$ and $\Gamma_2^+$ share the same character $-1$ under the mirror operation $M_{yz}$ in the $D_{2h}$ group, the overall character of Eq.~\ref{eq:AC_AC} is $(-1)^{|n|}$. This implies that when $n$ is even (odd), $\hat{H}^\prime|\psi_{i}\rangle$ is also even (odd), resulting in a nonzero (zero) matrix element $\mathcal{M}$ (see Case 1 in Note S2.3 of SM). Thus, the $n$=0 Floquet VB can be observed, while the spectral weights around the $\Gamma$ point for the $n$=1 sideband disappear (see Fig.~\ref{fig:2}B). 

For the ZZ ($p$-$pol.$) probe laser with the same AC ($s$-$pol.$) laser pumping as shown in Fig.~\ref{fig:2}C, the representation of $\hat{H}^\prime|\psi_{i}\rangle$ at the $\Gamma$ point is (see Case 2 in Note S2.3 of SM)
\begin{equation}
\label{eq:AC_ZZ}
(\underbrace{\Gamma_2^-\oplus\Gamma_4^-}_{probe})\otimes(\underbrace{\Gamma_3^-}_{pump})^{|n|}\otimes\underbrace{\Gamma_2^+}_{\left|\psi^v\right\rangle}
\end{equation}
The polarization direction of the probe laser is changed from that in Eq.~\ref{eq:AC_AC}, meaning that $\hat{H}^\prime|\psi_{i}\rangle$ transforms as an odd (even) function when $n$ is even (odd). This leads to the vanishing of the $n$=0 Floquet VB and the emergence of the $n$=1 sideband as a distinct dot-like feature around the $\Gamma$ point (see Fig.~\ref{fig:2}C). The Eq.~\ref{eq:AC_AC} and Eq.~\ref{eq:AC_ZZ}, derived purely from symmetry considerations, provide a general and analytical framework to comprehend the spectral features of Floquet sidebands. In particular, they reveal how the symmetry character of $\hat{H}^\prime|\psi_{i}\rangle$ alternates with the Floquet index $n$, enabling a direct, symmetry-based interpretation of TrARPES intensity plots.

Moreover, under the ZZ ($p$-$pol.$) pumping laser shown in Fig.~\ref{fig:2} (E and F), $\hat{H}^\prime|\psi_{i}\rangle$ is even (odd) under the AC (ZZ) probe laser, regardless of $n$ being odd or even (see Cases 3 and 4 in Note S2.3 of SM). This behavior stands in contrast to the AC ($s$-$pol.$) pumping case, where the symmetry character of $\hat{H}^\prime|\psi_i\rangle$ is sensitive to the Floquet index $n$. The key difference arises from the symmetry of the light-induced sideband wavefunction $|\psi_n^v\rangle$ under the mirror operation $M_{yz}$. For the ZZ ($p$-$pol.$) pumping, the vector potential $\mathbf{A}_{pu}$ lies in the scattering plane, rendering the term $(\mathbf{v(\mathbf{k}_{\|})}\cdot\mathbf{A}_{pu})$ in Eq.~\ref{eq:sideband} even under $M_{yz}$. As a result, $|\psi_n^v\rangle$ inherits the mirror symmetry of the original VB wavefunction $|\psi^v\rangle$, regardless of the Floquet index $n$. In contrast, under the AC ($s$-$pol.$) pumping laser, where $\mathbf{A}_{pu}$ is perpendicular to the scattering plane, the term $(\mathbf{v(\mathbf{k}_{\|})}\cdot\mathbf{A}_{pu})$ becomes odd under $M_{yz}$, leading to the character of $|\psi_n^v\rangle$ alternating with $(-1)^{|n|}$. This distinctive symmetry behavior accounts for the consistent appearance of sidebands around the $\Gamma$ point in Fig.~\ref{fig:2}E and their absence in Fig.~\ref{fig:2}F.

To systematically compare and highlight the symmetry properties of Floquet(-Volkov) sidebands under representative pump-probe configurations, we summarize the results of our group-theoretical analysis in Table~\ref{tab:sym}. This table presents a unified classification of the Floquet optical selection rules in monolayer BP, based on the interplay between the Floquet index $n$ and the symmetries of the equilibrium parent band, pumping and probe laser pulses. The classification not only recovers the known cases observed in specific experiments (e.g. Fig.~\ref{fig:2}C) but also provides predictive power for unexplored setups (e.g. Fig.~\ref{fig:2}F), serving as a guiding principle for future pump-probe spectroscopies.

\begin{figure}
	\centerline{\includegraphics[clip,width=0.7\linewidth]{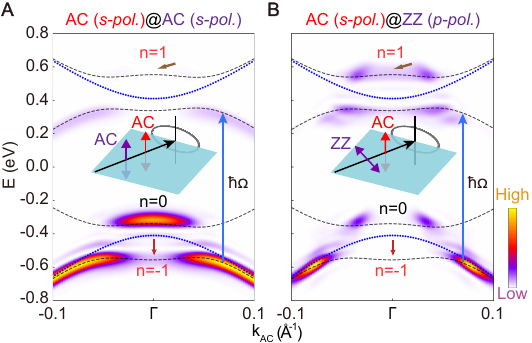}}
	\caption{\textbf{The simulated TrARPES intensity plots upon near-resonance pumping.} (\textbf{A}) The TrARPES intensity plot along the AC direction upon near-resonance pumping. The pumping and probe laser polarizations are along the AC ($s$-$pol.$) directions. The pump photon energy $\hbar\Omega$ is 0.88 eV and the peak intensity $I_{pu}$ is 4$\times10^{10}$ W/cm$^2$, while all other parameters are consistent with the below-gap pumping case. (\textbf{B}) Similar results to (A), but the polarization of the probe laser is along the ZZ ($p$-$pol.$) direction. The blue dot lines are equilibrium band structures and the black dashed lines are the Floquet band structures with the Floquet index $n$ as $\{-1,0,1\}$ for both panels. The red and brown arrows point to the regions of the $n$=$-$1 and $n$=1 Floquet CB and VB around the $\Gamma$ point, respectively.}
	\label{fig:3}
\end{figure}

To assess the universality of the established Floquet optical selection rules, we consider the near-resonance AC ($s$-$pol.$) pumping geometry \cite{zhou2023pseudospin}. Our TDDFT calculations reveal a strong band renormalization between the VB and CB induced by the resonant AC laser pumping (see~Fig.~\ref{fig:3}), which has been reported in previous experiments \cite{zhou2023pseudospin}. Despite the hybridization of the $n$=0 Floquet VB (CB) with other Floquet sidebands, their spectral distributions can still be interpreted by the Floquet optical selection rules. As an instance, the hybrid wavefunction $\left|\Psi_0^v\right\rangle$ of the $n$=0 Floquet VB at the $\Gamma$ point is
\begin{equation}
\label{eq:hybridization}
\left|\psi_i\right\rangle=\left|\Psi_0^v\right\rangle=a_0\left|\psi_0^v\right\rangle+b_{-1}\left|\psi_{-1}^c\right\rangle
\end{equation}
where $\left|\psi_0^v\right\rangle$ ($\left|\psi_{-1}^c\right\rangle$) denotes the wavefunction of the $n$=0 ($n$=$-$1) Floquet VB (CB) at the $\Gamma$ point without hybridization, $a_0$ and $b_{-1}$ are linear combination coefficients. Since both $\left|\psi_0^v\right\rangle$ and $\left|\psi_{-1}^c\right\rangle$ are odd upon the AC pumping (see Eq.~\ref{eq:sideband}), the hybrid wavefunction $\left|\Psi_0^v\right\rangle$ remains odd. This fact leads to $\hat{H}^\prime|\psi_i\rangle$ in Eq.~\ref{eq:matrix_element} being even (odd) under the AC (ZZ) probe laser. Consequently, the $n$=0 Floquet VB around the $\Gamma$ point retains a visible signature under the AC probe laser (see Fig.~\ref{fig:3}A), while its spectral weight is absent under the ZZ probe laser (see Fig.~\ref{fig:3}B). Similar features are also observed upon below-gap pumping: the $n$=0 Floquet VB around the $\Gamma$ point remains visible under the AC probe laser in Fig.~\ref{fig:2}B but is suppressed under the ZZ probe laser in Fig.~\ref{fig:2}C, consistent with the near-resonance pumping case under the same pump-probe configuration. Moreover, the red arrows in Fig.~\ref{fig:3} (A and B) highlight that the spectral features of the $n$=$-$1 Floquet CB resemble those of the $n$=0 Floquet VB around the $\Gamma$ point, as the hybrid wavefunction $|\Psi^c_{-1}\rangle$ of the $n$=$-$1 Floquet CB is also odd under the mirror operation $M_{yz}$ due to a similar linear combination of $\left|\psi_0^v\right\rangle$ and $\left|\psi_{-1}^c\right\rangle$.

Due to the establishment of a direct optical transition channel between the VB and CB, near-resonance pumping results in a slightly pumped-electron occupation around the CB edge. Similarly, the hybrid wavefunction $\left|\Psi_1^v\right\rangle$ of the $n$=1 Floquet VB at the $\Gamma$ point is also a linear combination of $\left|\psi_1^v\right\rangle$ and $\left|\psi_0^c\right\rangle$, exhibiting opposite symmetry and spectral properties to $\left|\Psi_0^v\right\rangle$ in Eq.~\ref{eq:hybridization}. As indicated by brown arrows in Fig.~\ref{fig:3} (A and B), while no signature is detected under the AC probe laser, the ZZ probe laser reveals these occupations. These facts indicate that even with the direct optical absorption between the VB and CB, as well as the hybridization between Floquet sidebands, the spectral weights observed in TrARPES still obey symmetry-governed Floquet optical selection rules, indicating their fundamental importance for the understanding of the Floquet phenomena.

\subsection*{Discussion}
From below-gap to near-resonance pumping, the Floquet optical selection rules consistently explain monolayer BP photoemission spectra, which share similar spectral characteristics with that of BP thin films. Moreover, we perform extensive TDDFT calculations by tuning the pump-probe delay time (figs. S3, S4 and S5 in SM), the peak intensity of the pumping laser (fig. S6 in SM), and the pulse duration of pumping and probe lasers (fig. S7 in SM) to capture the birth and death of Floquet(-Volkov) states, and we find the Floquet optical selection rules persist under different configurations. Although these selection rules are obtained from a group theory perspective without considering any scattering mechanism for these light-induced states, this framework can be effectively applied to interpret the monolayer BP photoemission spectra from our TDDFT simulations and experimental observations for the BP thin film. Furthermore, our paradigm developed in this work is not limited to BP. Still, it can be adaptable to momentum-resolved Floquet states for other materials, such as graphene, topological insulators, transition metal dichalcogenides, and hexagonal boron nitride under laser pumping (See Note S4 of SM). These results underline the pivotal role of symmetry for the coherent Floquet(-Volkov) states across a broader range of quantum materials.

At the same time, these findings also shed light on why the Floquet bands deserve focused attention within photoemission spectroscopy. Unlike equilibrium band structures, these Floquet states encode photon-dressed quantum coherence and dynamical symmetry properties that are intrinsically time-dependent and inaccessible through conventional ARPES. In contrast, TrARPES extends the capability of ARPES into the time domain, enabling direct observation of such non-equilibrium states with momentum and temporal resolution. When combined with symmetry-based Floquet optical selection rules, TrARPES offers a more refined level of control over light-matter interactions. By tuning the pump-probe geometry, one can selectively enhance or suppress spectral weights of specific photon-dressed states. This establishes a powerful approach not only for interpreting photoemission spectra in driven systems but also for tailoring quantum states on ultrafast timescales.

In summary, we propose general, symmetry-based Floquet optical selection rules, which intertwine the symmetries of the material, the pumping laser, and the probe laser within the Floquet framework. These rules provide qualitative and predictive insights into the spectral features of light-induced sidebands in TrARPES measurements. Using monolayer BP as a prototypical system, we validate these rules through fully \textit{ab initio} TDDFT simulations and demonstrate their explanatory power across different pump-probe configurations. Beyond reproducing previously known features, our simulations uncover distinct spectral weight distributions that are naturally explained within the proposed framework. While our analysis is taking BP as an example, the methodology is transferable to a broad class of quantum materials and experimental setups. Our theoretical framework provides a crucial tool for manipulating the symmetry of light-matter hybrid states in ultrafast spectroscopy, paving the way for advancements in symmetry-dominated ultrafast studies for more quantum materials.

\subsection*{Methods}

In the present work, all computational tasks were executed through the utilization of the real-space grid-based code, Octopus \cite{andrade2015real,tancogne2020octopus}. To commence, we elucidate the technical details for the ground-state calculations. The Kohn-Sham (KS) equations were solved iteratively to self-consistency with a relative convergence of the density as $10^{-7}$, and the grid spacing was set to 0.36 Bohr during the calculations. The adiabatic local density approximation (LDA) and Hartwigsen-Goedecker-Hutter (HGH) pseudopotentials \cite{hartwigsen1998relativistic} were utilized for the monolayer BP. Additionally, a $\Gamma$-centered $12\times10\times1$ k-grid in reciprocal space was employed, and the $z$ axis, orthogonal to the monolayer plane, was characterized using non-periodic boundaries with a real-space length of 120 Bohr. The lattice parameters along the $x$ and $y$ directions were 4.625 \angstrom{} and 3.299 \angstrom, respectively. The spin-orbit coupling (SOC) was not considered in the present analysis. The band gap of the monolayer BP was 0.82 eV in our calculations.

To simulate the pump-probe dynamics of the monolayer BP, we employed \emph{ab-initio} calculations based on the real-time TDDFT. During these calculations, we selected the dipole approximation and velocity gauge to describe the light-matter interaction. The dynamical evolution of the KS states was governed by the equation of motion in atomic units, which is expressed as:
\begin{equation}
i\partial_t\left|\Psi_{n\mathbf{k}}(t)\right\rangle=\left[\frac{1}{2}\left(-i\mathbf{\nabla}-\frac{\mathbf{A}(t)}{c}\right)^2+v_{KS}(t)\right]\left|\Psi_{n\mathbf{k}}(t)\right\rangle
\label{eq:td_ks_eq}
\end{equation}
Here, $\left|\Psi_{n\mathbf{k}}(t)\right\rangle$ represents the KS orbital with the momentum $\mathbf{k}$, $n$ is the band index, $c$ is the speed of light  (in atomic units), $v_{KS}(t)$ is the time-dependent KS potential, and $\mathbf{A}(t)$ is the total vector potential of the laser. To ensure convergence, the temporal spacing $\Delta t$ for solving the equation of motion in Eq.~\ref{eq:td_ks_eq} was chosen as $1.69\times10^{-3}$ fs. To describe the ARPES and TrARPES experiments, the laser vector potential $\mathbf{A}(t)$ in Eq.~\ref{eq:td_ks_eq} was divided into two terms:
\begin{equation}
    \mathbf{A}(t)=\mathbf{A}_{pu}(t)+\mathbf{A}_{pr}(t)
\end{equation}
Herein, $\mathbf{A}_{pu}(t)$ is the pumping laser to induce the sidebands and has the form:
\begin{equation}
\mathbf{A}_{pu}(t)=\mathbf{A}_{pu}\sin^2(\frac{\pi t}{T_{pu}})\cos(\Omega t)\Theta(T_{pu}-t)\Theta(t)
\label{eq:pump_laser}
\end{equation}
where $\mathbf{A}_{pu}=\frac{cE_{pu}}{\Omega}\boldsymbol{\epsilon}_{pu}$ with $E_{pu}$ as the electric field and $\boldsymbol{\epsilon}_{pu}$ as the polarization direction. $\Omega$ is the frequency and $T_{pu}$ is the total pulse duration of the pumping laser. $\Theta(t)$ is the time step function, ensuring that the pumping laser $\mathbf{A}_{pu}(t)$ is zero everywhere except for $t\in[0,T_{pu}]$.

$\mathbf{A}_{pr}(t)$ is the probe laser to detect the pumping laser-dressed system and is written as:
\begin{equation}
\mathbf{A}_{pr}(t)=\mathbf{A}_{pr}\sin^2(\frac{\pi t}{T_{pr}})\cos(\omega t)\Theta(T_{pr}-t)\Theta(t)
\label{eq:probe_laser}
\end{equation}
The physical meaning of the symbols in Eq.~\ref{eq:probe_laser} is similar to that in Eq.~\ref{eq:pump_laser}, but for the probe laser. Please note that the results presented in this work do not rely on the specific shape of the pumping and probe laser, so we have employed these particular forms for their convenience and simplicity.

In order to obtain the photoemission spectra of the monolayer BP under different pump-probe conditions, we employed the t-SURFFP \cite{tao2012photo,de2017first}
to evaluate photoemission probabilities from crystal surfaces without explicitly constructing continuum scattering states. The system is modeled as a slab geometry, which is periodic in-plane and finite along the out-of-plane direction.

To extract the photoelectron distribution, we computed the flux of the single-particle current through a surface $S$ placed 30 Bohr away from the slab surface, where complex absorbing potentials were also applied to prevent unphysical rescattering. The momentum-resolved photoelectron probability was then obtained from the modulus squared of the Volkov expansion coefficients as
\begin{equation}
\label{eq:Volkov_coe}
P(\mathbf{p}) = \lim_{t\rightarrow\infty}\frac{2}{N} \sum_{n=1}^{N/2} \left| b_n(\mathbf{p}, t) \right|^2
\end{equation}
where $\mathbf{p}$ is the total photoelectron momentum, $N$ is the number of escaped electrons and $b_n(\mathbf{p}, t)$ is the time-dependent expansion coefficient of the Kohn-Sham orbital $\left|\Psi_{n\mathbf{k}} (t)\right\rangle$ in Eq.~\ref{eq:td_ks_eq} onto Volkov states. The Volkov states are analytical solutions of the time-dependent Schr{\"o}dinger equation for a free electron in an external electromagnetic field. They provide a good approximation for photoelectrons in the vacuum and are given by
\begin{equation}
\chi_{\mathbf{p}}(\mathbf{r}, t) = \frac{1}{\sqrt{2\pi A}} e^{i \mathbf{p} \cdot \mathbf{r}} e^{i \Phi(\mathbf{p}, t)}
\end{equation}
where $A$ is the unit cell area and the Volkov phase $\Phi(\mathbf{p}, t)$ is defined as
\begin{equation}
\Phi(\mathbf{p}, t) = -\int_0^t \mathrm{d}\tau \left( \frac{1}{2} \left[ \mathbf{p} - \frac{\mathbf{A}(\tau)}{c} \right]^2 \right)
\end{equation}

The coefficient $b_n(\mathbf{p}, t)$ in Eq.~\ref{eq:Volkov_coe} was obtained by integrating the flux of the current projected onto plane waves over time:
\begin{equation}
b_n(\mathbf{p},t) = - \int_0^t \mathrm{d}\tau \oint_S \mathrm{d}\mathbf{s} \cdot \left\langle \chi_{\mathbf{p}}(\tau)\right|\hat{\mathbf{j}}(\tau)\left|\Psi_{n\mathbf{k}}(t)\right\rangle
\label{eq:tSURFF-final}
\end{equation}
where $\hat{\mathbf{j}}(t) = -i \nabla - \mathbf{A}(t)/c$ is the single-particle current operator, and the flux integral is taken over the surface $S$ with normal parallel to the non-periodic direction. Equation~\ref{eq:tSURFF-final} allows for efficient computation of the (Tr)ARPES signal using only the time-evolved wavefunctions evaluated on the surface $S$, where scattering states are well approximated by Volkov waves. This makes the method particularly well suited for systems with slab geometries.

As for the calculations of Floquet band structures, we carried out the non-interacting Floquet TDDFT approach merged in the Octopus code \cite{hubener2017creating}. This approach provides a quasi-static characterization of the pumping light-driven electronic states by performing a Floquet expansion of the TDDFT Hamiltonian, without explicitly including the probe pulse. An effective vector potential amplitude was used, corresponding to the value of the shaped pumping pulse at the temporal overlap with the probe pulse in the TrARPES simulations. Artificial doping of monolayer BP was also explored to visualize CB contributions in the ARPES intensity plot by adding 0.025 electronic charges per unit cell in the main text. We note that such slight doping has minimal effects on the band structures.

\nocite{li2014electrons,dresselhaus2007group,giovannini2020floquet}


\bibliography{Sci_Adv}
\bibliographystyle{sciencemag}


\paragraph*{Acknowledgments:}
We thank S.W. Jung and K.S. Kim for their helpful discussions.
\paragraph*{Funding:}
This work was supported by the National Natural Science Foundation of China (Grants No. 12234011 and No. 12374053) and the National Key Basic Research and Development Program of China (Grant No. 2024YFA1409100). B.F. and W.D. acknowledge the support of the Basic Science Center Project of NSFC (Grant No. 52388201), Innovation Program for Quantum Science and Technology (Grant No. 2023ZD0300500), and the Beijing Advanced Innovation Center for Future Chip (ICFC). U.D.G acknowledges support from the Marie Sk{\l}odowska-Curie Doctoral Network TIMES (Grant No. 101118915), SPARKLE (Grant No. 101169225), the Italian Ministry of University and Research (MUR) under the PRIN 2022 (Grant No. 2022PX279E$\_$003), and Next Generation EUPartenariato Esteso NQSTI - Spoke 2 (THENCE-PE00000023). A.R. acknowledges support from the Cluster of Excellence 'CUI: Advanced Imaging of Matter'- EXC 2056 - project ID 390715994, and Grupos Consolidados (IT1453-22). A.R. also acknowledges support from the Max Planck-New York City Center for Non-Equilibrium Quantum Phenomena. The Flatiron Institute is a division of the Simons Foundation. 
\paragraph*{Author contributions:}
P.T., A.R., and W.D. conceived the project. B.F. performed the TDDFT simulations with the help of U.D.G. and H.H. B.F. and P.T. analyzed the data. B.F. wrote the manuscript with input from U.D.G., H.H., S.Z., W.D., A.R., and P.T. All authors discussed the results and commented on the manuscript. 
\paragraph*{Competing interests:}
The authors declare that they have no competing interests.
\paragraph*{Data and materials availability:}
All data needed to evaluate the conclusions in the paper are present in the paper and/or the Supplementary Materials. 


\subsection*{Supplementary Materials}
This PDF file includes:\\
Notes S1 to S4\\
Tables S1 to S5\\
Figs. S1 to S8\\

\clearpage

\begin{table}
	\centering
	\scriptsize
	\caption{\textbf{Summary of symmetry analysis for four pump-probe configurations.} The group representations of $\hat{H}^\prime|\psi_n^v\rangle$ and resulting symmetry (even or odd) are listed for typical geometries.}
	\label{tab:sym}
	\begin{tabular}{ccccc}
		\hline \hline 
		& Fig.~\ref{fig:2}B, Fig.~\ref{fig:3}A & Fig.~\ref{fig:2}C, Fig.~\ref{fig:3}B & Fig.~\ref{fig:2}E & Fig.~\ref{fig:2}F \\ \hline
		Pump@Probe  & AC ($s$-$pol.$)@AC ($s$-$pol.$) & AC ($s$-$pol.$)@ZZ ($p$-$pol.$) &   ZZ ($p$-$pol.$)@AC ($s$-$pol.$) & ZZ ($p$-$pol.$)@ZZ ($p$-$pol.$) \\ \hline 
		\makecell{The group\\representation\\of $\hat{H}^\prime|\psi_n^v\rangle$}  &$\Gamma_3^-\otimes(\Gamma_3^-)^{|n|}\otimes\Gamma_2^+$ &$(\Gamma_2^-\oplus\Gamma_4^-)\otimes(\Gamma_3^-)^{|n|}\otimes\Gamma_2^+$ & $(\Gamma_3^-)\otimes(\Gamma_2^-\oplus\Gamma_4^-)^{|n|}\otimes\Gamma_2^+$ &$(\Gamma_2^-\oplus\Gamma_4^-)\otimes(\Gamma_2^-\oplus\Gamma_4^-)^{|n|}\otimes\Gamma_2^+$ \\ \hline 
		Symmetry &  \makecell{Even ($n$ is even)\\Odd ($n$ is odd)} & \makecell{Even ($n$ is odd)\\Odd ($n$ is even)} & Even (n is even $\&$ odd) & Odd (n is even $\&$ odd)  \\ \hline \hline
	\end{tabular}
\end{table}

\end{document}